\documentclass[aps,preprint,pra,showpacs]{revtex4}%
\usepackage{graphicx}
\usepackage{amssymb}
\usepackage{amsfonts}
\usepackage{amsmath}%
\usepackage{color}
\providecommand{\U}[1]{\protect\rule{.1in}{.1in}}
\def\be{\begin{equation}}
\def\ee{\end{equation}}
\def\bea{\begin{eqnarray}}
\def\eea{\end{eqnarray}}

\begin{document}
\title{Comment on ``Measurements without probabilities in the final state proposal''}
\author{Eliahu Cohen}
\affiliation{H.H. Wills Physics Laboratory, University of Bristol, Tyndall Avenue, Bristol
BS8 1TL, UK}
\author{Marcin Nowakowski}
\affiliation{Faculty of Applied Physics and Mathematics, ~Gdansk University of Technology, 80-952 Gdansk, Poland}
\affiliation{National Quantum Information Center of Gdansk, Andersa 27, 81-824 Sopot, Poland}

\date{\today}

\begin{abstract}
The final state proposal [G.T. Horowitz and J.M. Maldacena, J. High Energy Phys. {\bf 2004(2)}, 8 (2004)] is an attempt to relax the apparent tension between string theory and semiclassical arguments regarding the unitarity of black hole evaporation. The authors of [R. Bousso and D. Stanford, Phys. Rev. D {\bf  89}, 044038  (2014)] analyze thought experiments where an infalling observer first verifies the entanglement between early and late Hawking modes and then verifies the interior purification of the same Hawking particle. They claim that ``probabilities for outcomes of these measurements are not defined'' and therefore suggest that ``the final state proposal does not offer a consistent alternative to the firewall hypothesis.''
We show, in contrast, that one may define all the relevant probabilities based on the so-called ABL rule [Y. Aharonov, P.G. Bergmann, and J.L. Lebowitz, Phys. Rev. {\bf 134}, 1410 (1964)], which is better suited for this task than the decoherence functional. We thus assert that the analysis of Bousso and Stanford cannot yet rule out the final state proposal.

\end{abstract}

\maketitle

\section{Introduction}\label{sec1}

The discovery that black holes evaporate \cite{Haw1}, has led in the last decades to an intense debate. Semi-classical arguments, such as the original one due to Hawking \cite{Haw2}, suggest that during evaporation pure states evolve into mixed state and thus unitarity breaks. There is some evidence in string theory, however, that the evaporation of black holes should be unitary \cite{1,2,3,4}. As an approach for addressing this apparent contradiction, Horowitz and Maldacena (HM) suggested to impose a final boundary state at the black hole singularity \cite{HM}. This state entangles the infalling matter and infalling radiation, thus allowing teleportation of information outside the black hole via the outgoing Hawking modes. Several authors have further contributed to this approach, e.g. \cite{GP,LP,Ahn}. On the other hand, Almheiri, Marolf, Polchinski and Sully (AMPS) \cite{AMPS} pointed out
a fundamental conflict that arises in the description of the infalling observer who sees violation of entropy sub-additivity. Rather than breakdown of unitarity, they proposed as a resolution a singular ``firewall'' at the horizon.

Bousso and Stanford \cite{BS} have recently analyzed the final state proposal employing the AMPS scenario. They considered an infalling observer trying to assign a probability to a history with a pair of definite entanglement verifications - first between an early and a late Hawking outgoing modes $r_b$ and $b$, respectively and then between $b$ and its interior partner $\tilde{b}$. Based on their analysis, they concluded that such a probability does not exist. However, this analysis crucially depends on the decoherence functional formalism that was used.

In what follows, we will claim that the inexistence of probabilities is a direct result of the decoherent histories formalism, incorporating a strong consistent histories condition, rather than of the final state proposal. Hence, we believe that the latter cannot be excluded on these grounds. Moreover, following \cite{Poly} we show explicitly that the various probabilities can be calculated using the ABL rule \cite{ABL} within a different time-symmetric framework known as the Two-State Vector Formalism (TSVF) \cite{TSVF}.

\section{Two approaches, different notions of probability}\label{sec2}

To prepare the grounds for our conclusions regarding the final state proposal, we shall now briefly discuss the conceptual and quantitative differences between the consistent histories (CH) formalism and the TSVF.

\subsection{The Consistent Histories formalism}

The CH approach as an interpretation of quantum mechanics was introduced by Griffiths in 1984 \cite{Gri84}, and later discussed by Omn\'es \cite{Omn}. The decoherent histories approach due to Gell-Mann and Hartle \cite{GMH} is based on similar ideas. This approach is broadly compatible with standard quantum mechanics. However, the notion of measurement, through which probabilities are introduced in standard quantum theory, no longer plays a fundamental role. Instead, the time dependence of quantum systems is inherently stochastic, with probabilities given by the Born rule or its extensions.

The ordinary formula for transition probabilities in quantum mechanics is generalized to yield conditional probabilities for sequences of events at several different times, called ``consistent histories,'' via a criterion ensuring (with some limits) that classical rules for calculating probabilities, which are explicitly defined within the formalism, are satisfied. The resulting interpretive scheme applicable to closed quantum systems is explicitly time-symmetric and treats wave function collapse as a mere calculational tool \cite{Stanford}. It is important for the purposes of our Comment to note that the consistent histories formalism, mainly due to its projective tensor structure and consistency conditions, is a local theory.
%\textcolor{blue}{THE FOLLOWING STATEMENT IS A BIT AMBIGUOUS AND I'M AFRAID THE MEANING IS NOT CLEAR. CAN YOU PLEASE REPHRASE? and does not consider quantum %entanglement as such.}

%Furthermore, it makes no reference to processes of measurement and can be applied to sequences of microscopic or macroscopic events, or both, as long as the mathematical condition of consistency is satisfied.

This approach gives the same predictions as textbook quantum mechanics in the domain where the textbook rules can be properly applied, but in addition
allows a ``paradox-free'' discussion of microscopic properties and events. As will be clarified later, ``paradox-free'' importantly means in many cases avoiding an unambiguous prediction.

Within this formulation, classical mechanics emerges as a useful approximation of quantum mechanics under certain conditions. The price to be paid for this \cite{Stanford} is a set of rules for reasoning resembling, but also significantly different from, those which comprise quantum logic. An implication is the lack of a single universally-true state of affairs at each instant of time. However, there is a correspondence limit in which the new quantum logic becomes standard logic in the macroscopic world of everyday experience. The proposed quantum logic provided by the CH interpretation reduces to the familiar classical propositional logic in the same domain where classical mechanics serves as a good approximation to quantum mechanics.
Throughout the years, there have been many criticisms, e.g. \cite{Esp,Dow1,Dow2,Kent,Okon}, as well as replies \cite{Rep1,Rep2}.

Bousso and Stanford claim that the complications with probability assignment in the HM setup ``can be treated carefully
using the decoherence functional formalism''. We believe this is incorrect, but let us first outline their method as previously suggested by Gell-Mann and Hartle \cite{GMH}. The decoherence functional $D(\alpha,\alpha')$ depends on a pair of histories from an arbitrarily chosen family of histories, each described by a product of projection operators: $C_\alpha=\Pi_{\alpha_n}\Pi_{\alpha_{n-1}}...\Pi_{\alpha_1}$ and $C_\alpha'=\Pi_{\alpha'_n}\Pi_{\alpha'_{n-1}}...\Pi_{\alpha'_1}$. It is defined by:
\begin{equation}
D(\alpha,\alpha')=tr[\rho_fC_\alpha\rho_iC_\alpha'^\dagger],
\end{equation}
where $\rho_i$/$\rho_f$ are respectively the initial/final states of the system, normalized such that $tr[\rho_i\rho_f]=1$. Crucially, only when the decoherence functional is diagonal (in case of the strong consistency condition imposed on the accessible family of histories), probabilities can be assigned to the particular histories using the rule:
\begin{equation}\label{CHprob}
P(\alpha)=D(\alpha,\alpha).
\end{equation}
The purpose of the diagonality is ensuring consistency when calculating marginals:
\begin{equation}
\sum_\beta P(\alpha,\beta)=P(\alpha).
\end{equation}
Therefore, according to this approach, when the decoherence functional is not diagonal, probabilities cannot be  meaningfully assigned. Occasionally, the implementation of the measurement with the addition of an ancillary pointer helps to diagonalize the decoherene functional, but then the histories are changed as well. Both methods turn out to be problematic when applied to the scenario in \cite{BS} and we will focus on the first which can be most easily analyzed within the TSVF.

We note that there are different consistency conditions for the discussed decoherence functional $D(\alpha, \beta)$, including the weaker condition $D(\alpha, \beta)\approx \delta_{\alpha\beta} P(\alpha)$ (known as medium decoherence \cite{Gellmann} where $P(\alpha)$ stands for the probability of a history $H^{\alpha}$) or the linear positivity condition by Goldstein and Page \cite{Goldstein}. However, as Wilczek and Cotler observe \cite{WC1, WC2}, it is unclear at this moment if these variants are physically meaningful.
%THE FOLLOWING SENTENCE SEEMS UNRELATED TO THE ABOVE. CAN YOU RELOCATE IT OR REPHRASE? It is also helpful to assume normalization of histories with non-zero %weight which enables normalization of probability distributions for history events.

\subsection{The TSVF}

The sources of the TSVF date back to the 1964 paper \cite{ABL} by Aharonov, Bergman and Lebowitz (ABL) who derived a probability rule concerned with measurements performed on pre- and post-selected systems, i.e. systems with a final state specified in addition to the ordinary initial state. Given an initial state $|\psi_{i}\rangle$ and a final state $|\psi_{f}\rangle$ , the probability that an intermediate measurement of the non-degenerate operator $A=\sum_k a_kA_k$ characterized by the projectors $A_k$ yields the eigenvalue $a_{k}$ is
\begin{equation} \label{ABLrule}
Pr\left(a_{k}|\psi_{i},\psi_{f}\right)=\frac{\left|\langle \psi_f |A_k|\psi_i\rangle\right|^2}{\sum_j \left|\langle \psi_f |A_j|\psi_i\rangle\right|^2}.
\end{equation}
%If only an initial state is specified, Eq. \ref{ABLrule} reduces to the regular probability rule:
%\begin{equation} \label{rABL}
%Pr\left(a_{k}|\Psi_{i}\right)=\left|\left\langle a_{k}|\Psi_{i}\right\rangle \right|^{2}
%\end{equation}
%This can be obtained by summing over a complete set of final states, expressing the indifference to the final state.

The counterfactual use of this formula is controversial \cite{KComm,VReply}, but to the best of our knowledge it is widely accepted in cases where an actual measurement of $A$ is carried out, such as the case analyzed in \cite{BS}.

In subsequent works, the utility of the ABL rule was further understood. It was thus broadened to a new formulation of quantum mechanics - the TSVF \cite{TSVF} and a new interpretation of it - the Two-Time Interpretation \cite{TTIa,TTIb,TTIc}. The latter can be thought of as subtle kind of a hidden variables theory where the so-called measurement problem is solved when imposing a special boundary condition on the universe.

In contrast to the CH formalism and Eq. \ref{CHprob}, the ABL rule can be applied as long as the denominator is non-zero. This reflects the natural tendency of the TSVF to encounter all scenarios, even paradoxical ones, and provide unambiguous (although sometimes surprising) predictions. This formalism directly addresses cases where entanglement monogamy seems to be violated \cite{AC}, and was claimed to do so in a paradox-free manner \cite{Poly}.

The generalized form of the ABL rule which will be used next reads

\begin{equation} \label{ABLruleM}
Pr\left(k|\Pi_i,\Pi_f\right)=\frac{tr\left(P_k \Pi_i \rho\Pi_f \right)}{\sum_j tr\left(P_j \Pi_i \rho\Pi_f \right)},
\end{equation}
where $\rho$ is the state of the system, $\Pi_i$/$\Pi_f$ are projections on the initial/final states, respectively, and $P_k$ are the various projection operators that can be measured at any intermediate time.

\subsection{A recent discrepancy}

For a long time these two formulations of quantum mechanics coexisted peacefully. However, two recent papers \cite{OnLev,ReplyLev} have exposed and made accentuated the crucial differences between them. The consistent histories rules were built to avoid paradoxes when thinking classically about quantum
experiments \cite{ReplyLev}. Therefore the predictions of this formalism were shown to be different than those of the TSVF when a specific setup employing a nested-Mach-Zehnder interferometer was examined \cite{OnLev}. Moreover, the CH approach seems to capture less than the TSVF does in this specific experimental scenario with weak measurements \cite{OnLev,ReplyLev}. However, it seems that predictions of the Entangled Histories formalism \cite{WC3, Now} for the nested-Mach-Zehnder interferometer are in agreement with those of TSVF.

\section{Comparing the predictions of CH and TSVF} \label{sec3}

Let us start with the introductory example of Bousso and Stanford \cite{BS}:
A qubit is prepared with a definite spin along the $x$ direction, that is $\rho=|+\rangle\langle+|$. As correctly denoted by Bousso and Stanford, if we consider histories
that begin with this state, have definite values of the $z$ spin,
and then definite values of the $x$ spin again, we easily find (in case of no intermediate dynamics) that the
decoherence functional is not diagonal. Hence, according to this approach, probabilities for the
$z$ spin at intermediate times cannot be assigned. However, the ABL rule does allow to assign probabilities to spin measurements along the z-axis (denoted here by $\uparrow$ and $\downarrow$) during intermediate times, simply by calculating according to Eq. \ref{ABLrule}:
\begin{equation}
Pr\left(\uparrow|+,+\right)=\frac{\left|\langle + |\Pi_{\uparrow}|+\rangle\right|^2}{\left|\langle + |\Pi_{\uparrow}|+\rangle\right|^2+\left|\langle + |\Pi_{\downarrow}|+\rangle\right|^2}=1/2,
\end{equation}
where $\Pi_{\uparrow}$ and $\Pi_{\downarrow}$ are projectors onto the eigenspaces of $\uparrow$ and $\downarrow$ respectively. Similarly, $Pr\left(\downarrow|+,+\right)=1/2$.

Hence, we already see at this point that the ABL rule can unambiguously provide probabilities for measurement outcomes in a pre-/post-selected system, even in cases where the decoherence functional cannot do so. We shall use that now for analyzing the AMPS scenario within a post-selected model.

Let us repeat the details of the HM model, and within it the specific measurements which according to Bousso and Stanford have no probabilities. Similarly to the latter we shall denote by $b$ the Hawking quantum still in the near horizon zone, its interior
partner by $\tilde{b}$, forming together the infalling vacuum, and a
subsystem $r_b$ of the early Hawking radiation that purifies $b$
in the unitary out-state, with its interior partner $\tilde{r}_b$. The initial and final states of the black hole are:
\begin{equation} \label{rho_i}
\rho_i=|\Phi\rangle\langle\Phi|_{\tilde{r}_b,r_b}\otimes|\Phi\rangle\langle\Phi|_{\tilde{b},b}
\end{equation}
and
\begin{equation} \label{rho_f}
\rho_f=d^2|\Phi\rangle\langle\Phi|_{\tilde{r}_b,\tilde{b}}\otimes I_{b,r_b}
\end{equation}
respectively, where $I$ is the identity and $|\Phi\rangle_{x,y}$ is the maximally entangled state $|\Phi\rangle_{x,y}=\left(dim(x)\right)^{1/2}\sum_{i=1}^{dim(x)}|i\rangle|i\rangle$.
An observer now tries to assign probabilities to a history with definite $b$,$r_b$ result, followed by a definite $\tilde{b}$,$b$ result. In Table 1 of \cite{BS}, the $4\times4$ decoherence functional is calculated and shown to be non-diagonal. For instance, when the histories are $C_1\equiv \Pi_{\tilde{b},b}\Pi_{r_b,b}$ and $C_2 \equiv \left(1-\Pi_{\tilde{b},b}\right)\Pi_{r_b,b}$ the decoherence functional is: $D(1,2)=1/d^2-1/d^4 \neq 0$. Therefore, according to this set of assumptions, probabilities cannot be assigned to the various histories.

According to the ABL rule, however, these probabilities can be calculated straight-forwardly by applying Eq. \ref{ABLruleM}:
\begin{equation} \label{eq9}
Pr\left(1|\Pi_i,\Pi_f\right)=\frac{1}{3d^4-6d^2+4}
\end{equation}

\begin{equation}
Pr\left(2|\Pi_i,\Pi_f\right)=\frac{(d^2-1)^2}{3d^4-6d^2+4}
\end{equation}

\begin{equation}
Pr\left(3|\Pi_i,\Pi_f\right)=\frac{(d^2-1)^2}{3d^4-6d^2+4}
\end{equation}

\begin{equation} \label{eq12}
Pr\left(4|\Pi_i,\Pi_f\right)=\frac{(d^2-1)^2}{3d^4-6d^2+4},
\end{equation}

where $\Pi_i$ / $\Pi_f$ project on the initial/final states in Eq. \ref{rho_i} / Eq. \ref{rho_f}, respectively, and $k=1,2,3,4$ correspond to projections $P_k$ on histories $C_k$, when $C_1$ and $C_2$ were defined above and $C_3=\Pi_{\tilde{b},b}(1-\Pi_{r_b,b})$, $C_4=(1-\Pi_{\tilde{b},b})(1-\Pi_{r_b,b})$ (in accordance with the scenario in \cite{BS}).

 It can be easily seen that the probabilities in Eqs. \ref{eq9}-\ref{eq12} sum up to 1 as required, and the non-diagonal terms appearing in the CH approach do not play any role.

We stress again the one may doubt the application of the ABL rule when the decoherence functional fails to provide an unambiguous result, but as explained in \cite{ReplyLev}, it is very common for the former to have a greater explanatory power than the latter. In contrast to the decoherence functional, which carries with it some philosophical interpretations, the ABL rule is part and parcel of quantum mechanics and we do not see how one can deny its outcomes. In fact, we know that the specific type of measurements needed in this scenario is possible in principle without any violation of causality \cite{MPRQT} (see also \cite{BC_E}). Although the special post-selection implied by the final state proposal complicates this state of affairs, it does not imply cloning nor violation of monogamy, but rather a temporal product structure between events, which is allowed by quantum theory \cite{Poly}.

\section{Conclusion} \label{sec4}
To critically analyze a recent paper by Bousso and Stanford \cite{BS}, we compared in this Comment the CH with the TSVF approach. We have seen that while the assignment of probabilities within the first could be problematic in several cases, the second always allows to assign them. Therefore, the problem identified by Bousso and Stanford \cite{BS} when applying the CH approach to the final state proposal seems to originate from the shortcomings of this approach, and not from the proposal itself. It is worth mentioning that the TSVF approach naturally deals with apparent violations of entanglement monogamy \cite{Poly,AC} and provides interesting predictions regarding the values attained by $T_{\mu\nu}$ on the horizon \cite{Eng1,Eng2}. Furthermore, the Entangled Histories formalism \cite{Cot, Now}, allowing for a complex superposition of histories, reveals a non-local behavior in time and thus may overcome the setbacks of the CH approach. Unsurprisingly, it bears close similarly to the TSVF \cite{NC}.

\section*{Acknowledgements}

We wish to thank Yakir Aharonov, Jordan Cotler, Robert Griffiths, Nissan Itzhaki and Lev Vaidman for helpful discussions and remarks (although the content of this Comment does not necessarily reflect their views). E.C. was supported by ERC AdG NLST. M.N. was supported by a grant from the John Templeton Foundation. The opinions expressed in this publication are those of the authors and do not necessarily reflect the views of the John Templeton Foundation. Part of his work was performed at the National Quantum Information Center of Gdansk.

\end{document}